# Adaptive Fractional PID Controller for Robot Manipulator

H. Delavari[*,**], R. Ghaderi[*], A. Ranjbar N.[*], S.H. HosseinNia[***], S. Momani[****]

*Intelligent Research Group, Babol (Noushirvani) University of Technology,
Faculty of Electrical and Computer Engineering,
P.O. Box 47135-484, Babol, Iran,( r.ranjbar@nit.ac.ir)
**Hamedan university of Technology, Faculty of Electrical Engineering,(hdelavary@gmail.com),Hamedan, Iran
*** Department of Electrical, Electronic and Automation Engineering,
Industrial Engineering School,University of Extremadura, Badajoz, Spain
**** Department of Mathematics, Faculty of Science, University of Jordan,
Amman 11942, Jordan

**Abstract:** A Fractional adaptive PID (FPID) controller for a robot manipulator will be proposed. The PID parameters have been optimized by Genetic algorithm. The proposed controller is found robust by means of simulation in a tracking job. The validity of the proposed controller is shown by simulation of two-link robot manipulator. The result then is compared with integer type adaptive PID controller. It is found that when error signals in the learning stage are bounded, the trajectory of the robot converges to the desired one asymptotically.

## 1. INTRODUCTION

In recent years, the use of robotic arms in industrial applications has significantly been increased. Due to highly coupled nonlinear and time varying dynamic, the robot motion tracking control is one of the challenging problems. In addition uncertainty in the parameters of both mechanical part of manipulators and the actuating systems would cause more complexity. Many control algorithm such as computer torque method(Spong *et al.,1989*), optimal control(Green *et al.,2004*; Green *et al.,2001*), variable structure control (VSC) (Slotine *et al.,1991*), neural networks (NNs) (Barambones *et al.,2002*) and fuzzy system (Llama *et al.,2000*; Yoo *et al.,2000*; Cruz *et al.,2006*; Labiod *et al.,2005*) have been proposed to deal with this problem. Here because of nonlinear dynamics in robot manipulator, adaptive FPID has investigated.

Recently, fractional control of nonlinear systems begins to attract increasing attention to applications in control engineering. A fractional-order controller to stabilize an unstable open loop has been proposed in (Tavazoei *et al.,2007*). Adaptive fractional controller and adaptive PID controller are proposed to control chaotic systems in (Hosseinnia *et al.,2008*a) and (Hosseinnia *et al.,2008*b), respectively. An adaptive fractional PID controller has been studied in (Ladaci, *et al.,*2006). The performance of integer and fractional order controllers in a hexapod robot with joints at the legs having viscous friction and flexibility has been studied (SILVA *et al.,2004*). The experiments reveal that even when a good model of real operating dynamical phenomena exists, an implemented fractional-order $PD^\alpha$ controller would provide better robustness in comparison with an integer-order *PD* algorithm. Ref. (Valério *et al.,2003*) assesses relative merits and drawbacks of different digital implementations of non-integer order controllers for a robotic arm. A fractional derivative controller is used to shape an outer loop into a form of a fractional order integrator. The resultant step response behaves constant overshoot, independent of variations in the load, and robust, in a stability sense, to spillover effects. In (Duarte *et al.,2002*) several aspects of phenomena generated by a pseudo-inverse-based trajectory control of redundant manipulators are studied. Results are consistent which represent a step towards understanding the relation between chaotic phenomena and fractional calculus. Interactive communication between robot and human has been presented through a human interface design for path planning, using a fractional potential to take into account danger of obstacles in (Melchior *et al.,2002*).

In this paper, a novel adaptive FPID controller for *n-R* planar manipulator will be proposed. It must be noted that robot dynamics have integer order derivative and just the controller has fractional order derivative. Although FPID controller has good performance but in order to have more robustness, smoother control signal and smaller tracking error an adaptive FPID is proposed here.

It is known that the integration and differentiation can be specified in terms of the fundamental operator $_aD_t^\alpha$ known as differ-integration operator, where $a$ and $t$ are limited, and for the ordinary integration, $\alpha \in Z^+$ is the order of the operator, where $Z^+$ is the set of integers. In the case of non-integer integration and differ-integration, $\alpha \in R$. There are several basic definitions of the non-integer integration and differentiation. Among them, the most popular two definitions are the Grünwald–Letnikov (GL) and the Riemann–Liouville (RL) and the Caputo definitions. The GL definition is:

$$_aD_t^\alpha f(t) = \lim_{h \to 0} \frac{1}{h^\alpha} \sum_{j=0}^{[(t-\alpha)/h]} (-1)^j \binom{\alpha}{j} f(t-jh) \quad (1)$$

Where $[.]$ is a flooring-operator while the RL definition is given by:

$$_aD_t^\alpha f(t) = \frac{1}{\Gamma(n-\alpha)} \frac{d^n}{dt^n} \int_a^t \frac{f(\tau)}{(t-\tau)^{\alpha-n+1}} d\tau \quad (2)$$

For $(n-1 < \alpha < n)$ and $\Gamma(x)$ is the well-known Euler's Gamma function. According to recent literatures, an equation contains fractional-order derivative/integration is referred to as fractional-order ordinary differential/integral equation (FO-ODE for the former in short). FO-ODE is defined as an equation containing fractional-order derivatives, which is as follows:

$$g(t, x, {_aD_t^{\alpha_1}}x, {_aD_t^{\alpha_2}}x,...) = 0 \quad (3)$$

where $\alpha_k \in R^+$. The Caputo's definition can be written as

$$_aD_t^\alpha f(t) = \frac{1}{\Gamma(\alpha-n)} \int_a^t \frac{f^{(n)}(\tau)}{(t-\tau)^{\alpha-n+1}} d\tau \quad (4)$$

For $(n-1 < \alpha < n)$. In this paper, fractional operators are implemented by continuous CRONE approximations. In the robotics context, it is known that PID controllers are sensitive to uncertainties which arise from imprecise knowledge of the kinematics and dynamics, and from joint and link flexibility, actuator dynamics, friction, sensor noise, and unknown loads, then an adaptive fractional PID (FPID) controller is proposed to use the robustness of fractional order controller and adaptive controller. Three PID control gains in (10), $K_p$, $K_i$, and $K_d$, are adjustable parameters and will be updated online with an adequate adaptation mechanism and the $\lambda$ and $\mu$ will be determined offline, here a genetic algorithm will be used. By introducing a supervisory controller, the stability of the closed-loop FPID control system under external disturbance can be guaranteed. In previous works, FPID controller parameters $K_p$, $K_i$, and $K_d$ are constant during the control process but here in this paper these parameters will be updated online with an adequate adaptation mechanism to have better results.

This paper is organized as follows: Section 2 presents dynamic of $n$-R planar manipulator. Adaptive FPID controller designation and tuning FPID controller and the effect of fractional parameters are discussed in Section 3 and Section 4, respectively. Simulation Result is studied in section 5. Finally, concluding remarks are drawn in Section 6.

## 2. DYNAMICS OF $n$-R PLANAR MANIPULATOR

For controller design, it is necessary to have a mathematical model of system. A robot manipulator can be defined as an open kinematics chain of usually rigid links in Fig.1. According to the Lagrangian formulation, dynamic of an $n$-joint robot manipulator with revolute joints can be formulated as:

$$M(q)\ddot{q} + C(q,\dot{q})\dot{q} + G(q) = U \quad (5)$$

where $q, \dot{q}, \ddot{q} \in R^n$ are $n$-**D** joint variables and $U$ represents $n$-**D** generalized forces. $M(q) \in R^{n \times n}$ is a symmetric and positive definite inertia matrix, $C(q,\dot{q})\dot{q}$, the Coriolis /centripetal vector, and $G(q)$ is the gravity vector. A robot manipulator generally presents uncertainties such as frictions and disturbances. Therefore, (5) can be written as:

$$M(q)\ddot{q} + C(q,\dot{q})\dot{q} + G(q) + D_{ist} = U \quad (6)$$

where $D_{ist}$ stands for uncertainty of the dynamic, including $F(q)$ frictions and disturbance. The friction in dynamic (6) is of the form $F(q) = F_v\dot{q} + F_d \text{sgn}(\dot{q})$ where $F_v$, $F_d$ are the coefficient matrix of viscous friction and dynamic friction term, respectively.

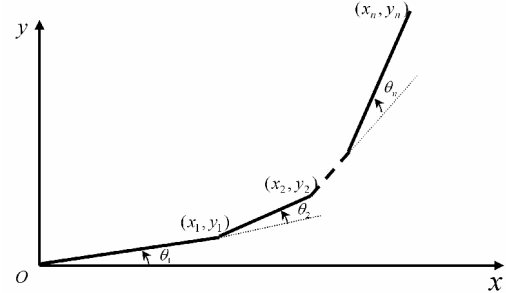

Fig. (1): $n$-link planar robot

## 3. DESIGN OF FRACTIONAL ADAPTIVE PID CONTROLLER

Let us consider a *2*-degree of freedom dynamic of robot manipulators as:

$$\ddot{q} = -(M(q))^{-1}[C(q,\dot{q})\dot{q} + G(q) + D_{ist}] + (M(q))^{-1}U \quad (7)$$

where $M(q) \in R^{2\times2}$, $C(q,\dot{q}) \in R^{2\times2}$, $G(q) \in R^{2\times1}$, $D_{ist} \in R^{2\times1}$ and $U \in R^{2\times1}$. Suppose that we can choose a gain vector $K = [k_0, k_1]^T$ such that roots of $S^2 + k_1 S + k_0 = 0$ are in the open left-half complex plane, and let $q_d$ is desired state and $E = [e \; \dot{e}]^T$ with $e = q_d - q$ is the error vector signal between the desired and actual outputs. Define $Q_d = [q_d \; \dot{q}_d]^T$ and assume that both $q_d$ and $\dot{q}_d$ are bounded, i.e., $\|Q_d\|_\infty = \sup_{t \geq 0} \|Q_d(t)\| < \infty$. Now let a feedback linearization controller be given by:

$$u^* = C(q,\dot{q})\dot{q} + G(q) + D_{ist} + M(q)(K^T E + \ddot{q}_d) \quad (8)$$

Substituting (4) into (2), we have $\ddot{e} + k_1\dot{e} + k_0 = 0$ it means that we have $e(t) \to 0$ as $t \to \infty$ i.e. $q \to q_d$ asymptotically. It must be noted that we assume the constraint set $\Psi_x = \{X \in R^2 : \|X\| \leq M_x\}$ where $M_x$ is a pre-specified parameter. It is assumed that the state trajectory of system $X$ never reaches the boundary $\Psi_x$ during the control procedure, and we define that $\|q_d\|_\infty \leq M_x$. Now we choose the control input by:

$$U = u_{FPID} + u_s \quad (9)$$

A control input $U \in R^{2\times1}$ will be generated from a FPID and a supervisory controller. The latter will be designed to provide the stability. The schematic diagram of control process is shown in Fig.2. The fractional controller part i.e. $u_{FPID}$ will be generated according to:

$$u_{FPID} = K_p^T diag(e(t)) + K_i^T D^{-\lambda} diag(e(t)) + K_d^T D^{\mu} diag(e(t)) \qquad (10)$$

Where, $K_p \in R^{2 \times 1}$, $K_i \in R^{2 \times 1}$ and $K_d \in R^{2 \times 1}$ whilst $e(t) \in R^{2 \times 1}$ stands for the error vector and $\lambda$, $\mu$ is fractional order integral and differential.

Substituting the control input into equation (7), we have, in view of (8):

$$\ddot{q} = -(M(q))^{-1}[C(q,\dot{q})\dot{q} + G(q) + D_{ist}]$$
$$+ (M(q))^{-1}(u_{FPID} + u_s)$$
$$= -(M(q))^{-1}[C(q,\dot{q})\dot{q} + G(q) + D_{ist}] \qquad (11)$$
$$+ (M(q))^{-1}(u_{FPID} + u_s + u^* - u^*)$$
$$= K^T E + \ddot{q}_d + (M(q))^{-1}(u_{FPID} + u_s - u^*)$$

With the error vector signal $e = q_d - q$ one can obtain:

$$\ddot{e} = -K^T E - (M(q))^{-1}(u_{FPID} + u_s - u^*) \qquad (12)$$

where $E = [e, \dot{e}]^T$ and $K = [k_0, k_1]^T$. Defining:

$$A = \begin{bmatrix} 0 & 1 \\ -k_0 & -k_1 \end{bmatrix}, B = \begin{bmatrix} 0 \\ 1 \end{bmatrix}$$

Changes equation (12) into a companion form of:

$$\dot{E} = AE + (M(q))^{-1} B(-u_{FPID} - u_s + u^*) \qquad (13)$$

The following function will be a candidate of the Lyapunov:

$$V_u = (1/2) E^T PE \qquad (14)$$

where $P$ is a positive definite matrix which will be obtained from the following relation:

$$A^T P + PA = -Q \qquad (15)$$

where $Q$ is a positive definite matrix.

This may be designed by the user in advance. Let us Define $V_M$ as:

$$V_M = (1/2) \min(\lambda(P))(M_x - \|q_d\|_\infty)^2 \qquad (16)$$

Leads us to the following equation:

$$V_u = (1/2) \min(\lambda(P))(\|E\|)^2 \geq$$
$$(1/2) \min(\lambda(P))(\|q\| - \|q_d\|)^2 \geq \qquad (17)$$
$$(1/2) \min(\lambda(P))(M_x - \|q_d\|_\infty)^2 = V_M$$

Where $\min(\lambda(P))$ denotes the minimum value of eigenvalue of matrix P. For the case $V_u < V_M$ one obtains $\|q_d\| < M_x$. Meanwhile when $\dot{V}_u$ satisfies in equation (14), leads us to:

$$\dot{V}_u = (1/2) E^T (A^T P + PA) E +$$
$$EP(M(q))^{-1} B(-u_{FPID} - u_s + u^*) =$$
$$-(1/2) E^T QE + E^T P(M(q))^{-1} B(-u_{FPID} - u_s + u^*) \qquad (18)$$
$$\leq -(1/2) E^T QE + |E^T P(M(q))^{-1} B|(|u_{FPID}| + |u^*|)$$
$$- E^T P(M(q))^{-1} B u_s$$

If $|G| \leq G_u$ and $|D_{ist}| \leq d_u$ therefore:

$$u^* \leq G_u + d_u + |C(q,\dot{q})\dot{q}| + |M(q)K^T E| + |M(q)\ddot{q}_d| \qquad (19)$$

Selecting a supervisory controller:

$$u_s = \begin{cases} 0 & ,V_u < V_M \\ sgn(E^T P(M(q))^{-1} B)\{G_u + d_u + |C(q,\dot{q})\dot{q}| & \\ + |M(q)K^T E| + |M(q)\ddot{q}_d| + |u_{FPID}|\} & ,V_u \geq V_M \end{cases} \qquad (20)$$

Together with (18) guarantees the stability.

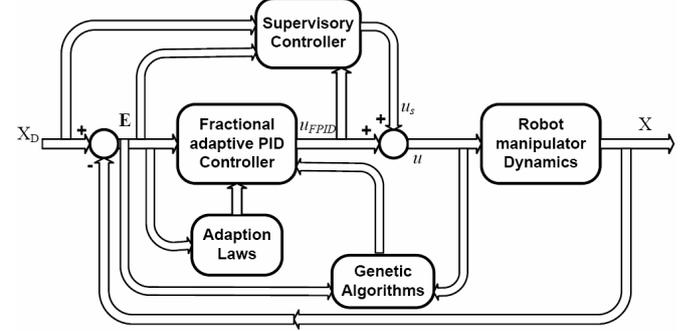

Fig. 2. Robust fractional adaptive PID control for robot manipulator system.

## 4. FPID CONTROLLER TUNING AND PARAMETERS OF THE FRACTIONAL CONTROLLER

When a supervisory controller is designed, parameters of the FPID controller i.e. $\lambda$ and $\mu$ have to be determined, here a genetic algorithm has used to determine these parameters. An adaptive technique based on gradient method can be used to tune (determine) parameters of FPID controller coefficients ($K_P, K_I, K_D$).

*4.1 Tuning FPID controller based on adaptive law*

In order to tune the parameters, an adaptation mechanism should be designed. Let us define a designed signal $q_r$ as:

$$\dot{q}_r = \ddot{q}_d + K^T E \qquad (21)$$

This may be achieved using a sliding mode controller. A sliding surface for each arm can be defined as:

$$S = \dot{q} - q_r \qquad (22)$$

When the system settles at the sliding mode, i.e. $S = 0$ therefore:

$$\dot{q} = q_r \qquad (23)$$

Replacement of relation (23) in (21) approaches:

$$\ddot{q} = \ddot{q}_d + K^T E \qquad (24)$$

Or in a scalar format:

$$\ddot{e} + k_1 \dot{e} + k_0 e = 0 \qquad (25)$$

The error approaches zero when $k_0$ and $k_1$ are assigned appropriately. To investigate the stability a Lyapunov function will be candidate as follows:

$$V_s = (1/2) S^2 > 0 \qquad (26)$$

In order to stay in the sliding mode the differentiation of the Lyapunov function must be negative definite. Satisfying the necessity in (27) guarantees states to keep staying in the surface $S = 0$.

$$\dot{V}_s = S\dot{S} < 0 \tag{27}$$

For a proper adaptation mechanism, the gradient method will be used to condition in (27). The gradient search algorithm is calculated in the opposite direction of the energy flow. The derivative of the sliding surface is:

$$\dot{S} = \ddot{q} - \dot{q}_r = [-(M(q))^{-1}(C\dot{q} - G - D_{ist}) + (M(q))^{-1}(u_{FPID} + u_s) - \dot{q}_r] \tag{28}$$

Multiplication of $S$ to both sides of equation (28) yields:

$$S\dot{S} = S.[-(M(q))^{-1}(C\dot{q} - G - D_{ist}) + (M(q))^{-1}(u_{FPID} + u_s) - \dot{q}_r] \tag{29}$$

Accordingly, the PID coefficients will be updated as:

$$\dot{K}_P = -grad(S\dot{S}, K_p)\gamma = -grad(S\dot{S}, u_{FPID}).grad(u_{FPID}, K_p).\gamma = -Se(t)^T \gamma \tag{30}$$

$$\dot{K}_I = -grad(S\dot{S}, K_i)\gamma = -grad(S\dot{S}, u_{FPID}).grad(u_{FPID}, K_i).\gamma = -S\left[\int_0^t e(\tau)d\tau\right]^T \gamma \tag{31}$$

$$\dot{K}_d = -grad(S\dot{S}, K_d)\gamma = -grad(S\dot{S}, u_{FPID}).grad(u_{FPID}, K_d).\gamma = -S\left[\frac{d}{dt}e(t)\right]^T .\gamma \tag{32}$$

where $\gamma \in R^{2\times 1}$ is a positive learning coefficients vector. As proposed, best fit parameters of fractional controller will be updated using Genetic Algorithm. It is of the goal to verify the performance of an artificial FPID with respect to the classic controller.

*4.2 Genetic Algorithm (GA) based Tuning of fractional parameters*

The following cost function will be defined to simultaneously minimize the control energy (effort) and the tracking error.

$$I = \int w_1(q - q_d)^2 + w_2 u^2 \tag{33}$$

where $u^2$ is the control energy, $w_i, i=1,2$ is the weighting coefficients, $q$ and $q_d$ are the system output and the desired one, respectively. The search space to find FPID coefficient is a real number. The parameters of the GA with trial and error are chosen as: Population size = *60,* Crossover probability = *0.75,* Generations = *80,* Mutation probability = *0.03*. In this method Genetic algorithm is used offline. The outcome of Genetic Algorithm verifies that FPID achieves less cost function.

## 5. SIMULATION RESULTS

*5.1 A two-link robot manipulator*

In this section, the proposed adaptive FPID controller is used on a two-link robot manipulator in Fig.3, where parameter matrices are given by:

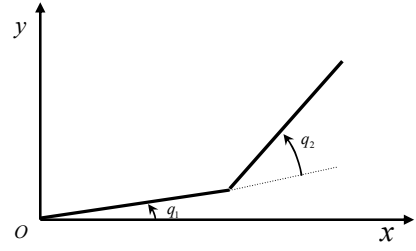

Fig. 3: A two-link robot manipulator

$$\begin{bmatrix} M_{11} & M_{12} \\ M_{21} & M_{22} \end{bmatrix}\ddot{q} + \begin{bmatrix} C_{11} & C_{12} \\ C_{21} & C_{22} \end{bmatrix}\dot{q} + \begin{bmatrix} G_1 \\ G_2 \end{bmatrix} + \begin{bmatrix} D_{ist1} \\ D_{ist2} \end{bmatrix} = \begin{bmatrix} U_1 \\ U_2 \end{bmatrix} \tag{34}$$

where:

$$\begin{aligned}
M_{11} &= (m_1 + m_2)l_1^2 + m_2 l_2^2 + 2m_2 l_1 l_2 \cos q_2, \\
M_{12} &= m_2 l_2^2 + m_2 l_1 l_2 \cos q_2, M_{21} = M_{12}, \\
M_{22} &= m_2 l_2^2, C_{11} = -2(m_2 + m_3)l_1 l_2 \dot{q}_2 \sin q_2 \\
C_{22} &= 0, C_{12} = -(m_2 + m_3)l_1 l_2 \dot{q}_2 \sin q_2, \\
C_{21} &= (m_2 + m_3)l_1 l_2 \dot{q}_1 \sin q_2 \\
G_1 &= g(m_1 l_1 \cos q_1 + m_2 l_1 \cos q_1 + m_2 l_2 \cos(q_1 + q_2)) \\
G_2 &= g m_2 l_2 \cos(q_1 + q_2), D_{ist\,k} = -d_k(t),\ k=1,2 \\
U_1 &= \tau_1, U_2 = \tau_2
\end{aligned} \tag{35}$$

Where $m_1$ and $m_2$ are masses, $l_1$ and $l_2$ are lengths of link *1* and *2*, respectively. $\tau_1$ and $\tau_2$ are driving force and torque at joints *1* and *2* respectively. $d_k(t)$ is unknown but bounded external disturbance with:

$$|d_k(t)| \le d_{uk},\ k=1,2 \tag{36}$$

During the simulation, parameters are chosen as: $m_1 = 1kg$, $m_2 = 1kg$, $l_1 = 1m$, $l_2 = 0.8m$, $g = 9.8 m/s^2$ and the initial conditions are $[q_1(0), \dot{q}_1(0), q_2(0), \dot{q}_2(0)]^T = [0, -0.2, 0.35, 0.5]^T$. Considering the disturbance $d_1(t) = 0.4\cos(5\pi t)\,Nm$ and $d_2(t) = 0.6\cos(5\pi t)\,Nm$, an upper limit of related functions are assumed as follows:

$$|D_{ist}(.)| = \begin{bmatrix} |0.4\cos(5\pi t)| \\ |0.6\cos(5\pi t)| \end{bmatrix} \le \begin{bmatrix} 0.4 \\ 0.6 \end{bmatrix} \tag{37}$$

Primary setting of FPID coefficients are chosen as: $k_{P2}(0) = 10$, $k_{I1}(0) = 25$, $K_{D1}(0) = 20$ $k_{P2}(0) = 12$ $k_{I2}(0) = 20$, $K_{D2}(0) = 15$ whilst the learning rate has been selected as $\gamma_1 = \gamma_2 = 25$. GA will be used to minimize the cost in (33) to find an optimum value of the fractional controller. Meanwhile, the search space of $\mu$ and $\lambda$ is a real space of $[0,1]$ and $[-1,0]$ respectively. Parameters are averaged over some different run and found as: $\lambda_1 = 0.901$, $\mu_1 = -0.921$, $\lambda_2 = 0.931$, and $\mu_2 = -0.932$.

The robot manipulator joints are driven according to the following desired trajectory:

$$q_{1d} = \sin(t) \tag{38}$$

$$q_{2d} = \sin(t)$$

The controller parameter $k_{11} = k_{12}$ and $k_{21} = k_{22}$ have been assigned as 1 and 2 respectively. The performance is shown improved (Fig. 4- Fig. 5) when the search space is expanded from the integer type to the fractional one. FPID controller reduces the tracking error significantly.

In comparison with equation (15) a matrix gain $A_1, A_2$ can be found and choosing $Q_1, Q_2$ as:

$$A_1 = A_2 = \begin{bmatrix} 0 & 1 \\ -0.5 & -1 \end{bmatrix}, Q_1 = Q_2 = \begin{bmatrix} 1 & 0 \\ 0 & 1 \end{bmatrix} \qquad (39)$$

Regarding equations (15), yields the symmetric and definite positive matrix $p$ as: $P_1 = P_2 = \begin{bmatrix} 1.75 & 1 \\ 1 & 1.5 \end{bmatrix}$. Tracking of $q_1, q_{1d}$ and $q_2, q_{2d}$ by using integer order and fractional order controller have been shown in Fig.4.$a$, Fig.5.$a$ respectively, whereas control signals are shown in Fig. 4.$b$, and Fig.5.$b$. As it can be seen the tracking is made enhanced together with more feasibility of the control signal, if the fractional controller is used.

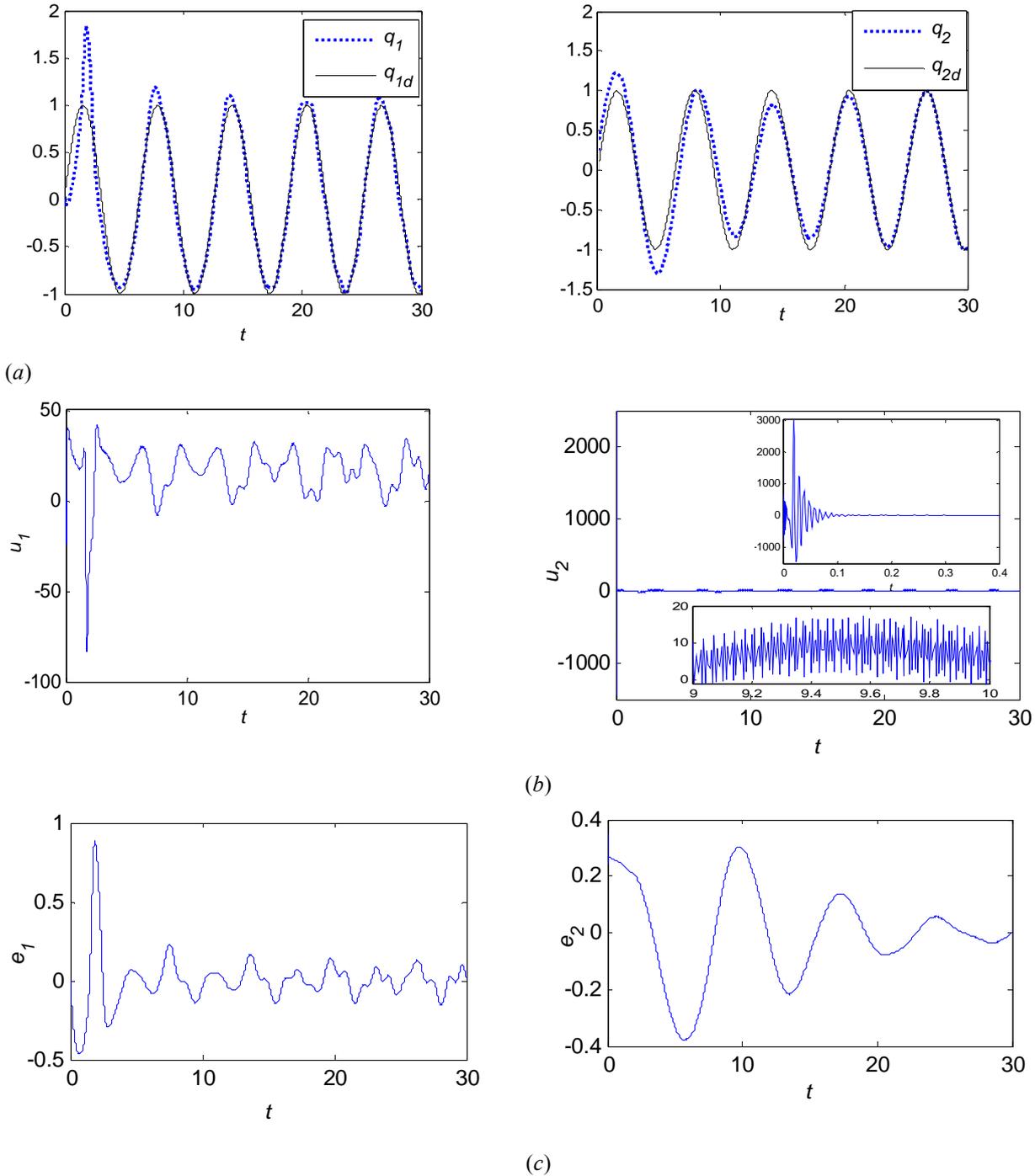

Fig. 4: Simulation result of an integer-order controller, ($a$) Tracking of desired and actual state, ($b$) control signals, ($c$) Error of tracking

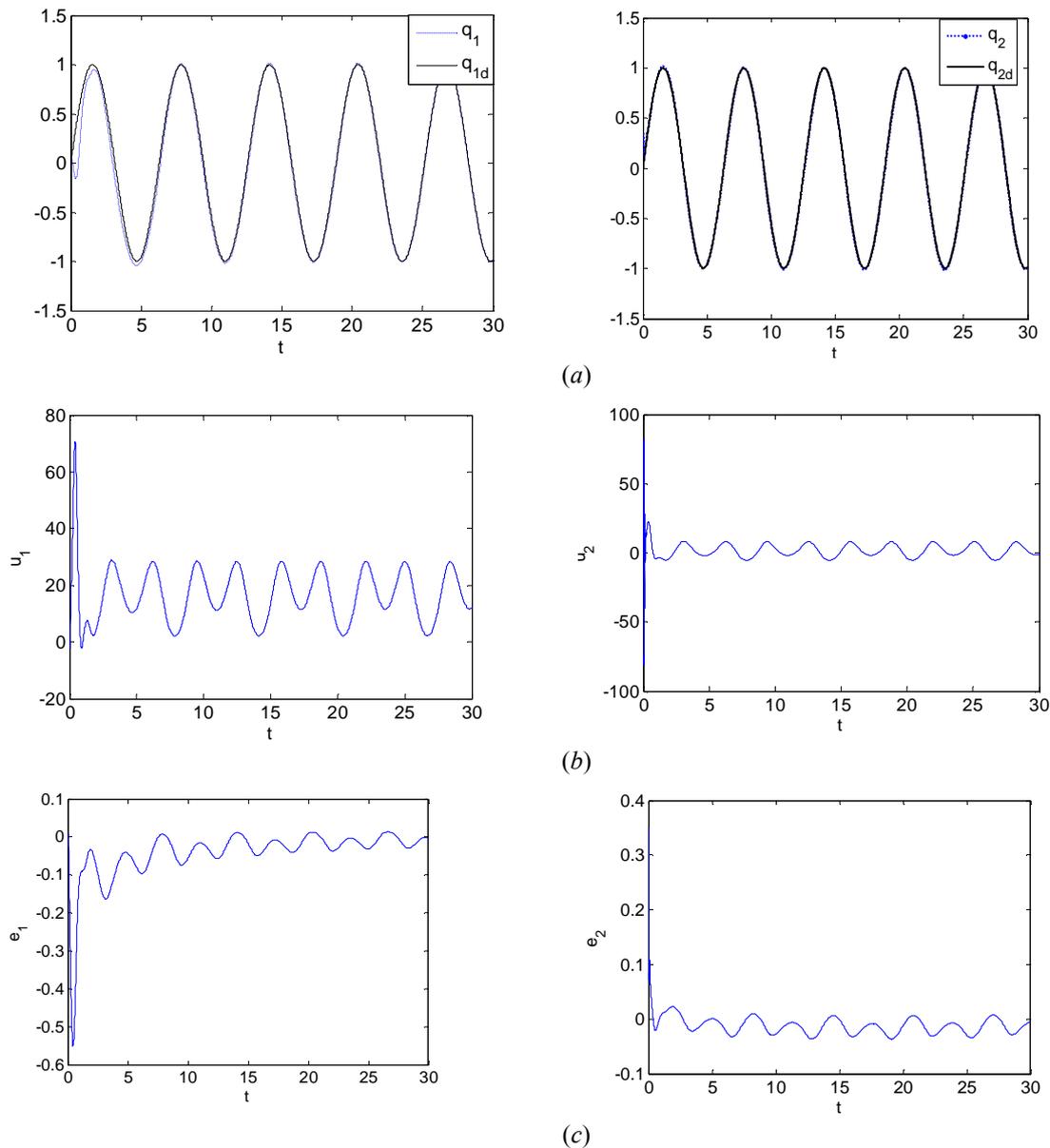

Fig. 5: Simulation result of a fractional-order controller, (*a*) Tracking of desired and actual state, (*b*) control signals, (*c*) Error of tracking

## 6. CONCLUSION

A trajectory control of a robot manipulator using adaptive FPID controller and integer order PID control algorithms has been studied. Three PID control gains in (10), $K_p$, $K_i$, and $K_d$, are adjustable parameters and will be updated online with an adequate adaptation mechanism and the $\lambda$ and $\mu$ will be determined offline, here a genetic algorithm will used. The stability of closed-loop FPID control system can be guaranteed under external disturbance by using the Lyapunov approach with a supervisory controller. The proposed control technique was analyzed for two types of manipulators having different dynamic (It is clear that robot dynamics have integer order derivative and just the controller has fractional order derivative). A genetic algorithm has also been used to choose the order of the fractional order. Simulation result verifies the significance of the fractional controller.